\documentclass[aps,prl,twocolumn,groupedaddress,showpacs,tighten]{revtex4}

\usepackage{graphicx}
\usepackage{times}

\newcommand{\beq}{\begin{equation}}
\newcommand{\eeq}{\end{equation}}
\newcommand{\msun}{{\rm M_\odot}}
\newcommand{\lsun}{{\rm L_\odot}}
\newcommand{\GeV}{{\rm GeV}}
\newcommand{\cm}{{\rm cm}}
\newcommand{\bh}{{\bullet}}
\newcommand{\pc}{{\rm pc}}
\newcommand{\kpc}{{\rm kpc}}
\newcommand{\kms}{\ {\rm km}\,{\rm s}^{-1}}

\def\kpc{\ {\rm kpc}}

\begin{document}
\title{Mini-dark halos with intermediate mass black holes}
\author{HongSheng Zhao and Joseph Silk}
\date{1 June 2005 on Phys. Rev. Letters 95, 011301}
\begin{abstract}
We argue that the Milky Way (MW) contains thousands intermediate mass black holes (IMBHs) 
and minihalos with a fraction of IMBHs still being enshrouded in extremely dense mini-spikes 
of dark matter (DM) particles.  
Each containing $10^6\msun$ of dark particles and no baryons in a sphere of 50 pc radius, 
the minihalos are dense enough to survive the Milky Way tide
with the nearest minihalo within 2 kpc from the Sun.  
The IMBH is formed off-centre in a minihalo by gas accretion, and 
its growth adiabatically compresses a finite density of surrounding dark matter
into a $r^{-1.5}$ mini-spike.  Some IMBHs recentre on their minihalos 
after dynamical friction, and some IMBHs are ejected by birth kicks.  
Detectable by GLAST, the mini-spikes and minihalos should stand out the background and 
dominate the neutralino annihilation in the smooth MW and satellite galaxies.
If they are the unidentified EGRET sources, upper limits can be set on
the branching ratio of neutralino annihilation.
The supermassive BH of the MW, if orginates from an IMBH, 
is also likely enshrouded with a mini-spike.
\end{abstract}

%\begin{keywords}
%galaxies: halos -- galaxies: interactions -- cosmology: theory --dark matter.
%\end{keywords}

\maketitle
%\narrowtext

Supermassive black holes (SMBH) likely grow from smaller intermediate mass black holes (IMBH)
by gas accretion; to be massive enough and early enough, a seed of $\sim 1000 M_\odot$ 
at $z \sim 20$ is necessary to grow to a $3 \times 10^9\msun$ SMBH at $z=6$ via Eddington accretion of
gas (e.g., \cite{MarelReview,GnedinTiming} and references therein). 
Gas accretion is favoured because of the empirical result \cite{Soltan82,YT02} 
that requires radiatively efficient accretion in order to
account for both the quasar luminosity function and the mass in black holes
at present.  Conversely, once the quasar/AGN phase is over, black holes
undergo radiatively inefficient accretion as is the case for Sgr A$^\ast$
(e.g., \cite{Goldston05}).  Here we explore
implications of growth of SMBH by accretion in the protogalactic environment.
We will study the evolution of the seed IMBHs.  These smaller IMBHs are
plausibly formed in a subset of the first objects, where accretion rates are
high.  Hierarchical merging of minihalos leads to a population of IMBHs in
MW-sized halos.  A mini-spike of CDM is inevitable around the IMBH due
to adiabatic growth in the minihalo. The mini-spike is robust to halo
merging, and survives around IMBHs that populate the halo.  The subsequent
role of BH merging depends on dynamical friction and loss cone refilling.
These processes are likely to 
nurture the growth of the central seed IMBH  into a supermassive BH.

The IMBHs, and in particular, the dark matter spikes, are
potentially observable via neutralino annihilations. We mention  
two specific examples. Firstly, a central  IMBH is inferred
to be  associated with a group
of massive stars IRS 13 offset by about 2 arcsec from the Galactic
centre (SgrA*) \cite{Maillard04}. Secondly, 
the Whipple/HESS detection of  $>3$TeV gamma rays from the
Galactic centre (\cite{Referee1,Referee2,Referee3}) 
would require a high annihilation cross section $\sim$10 pb if
this emission is due to a neutralino annihilation signal produced by an NFW\cite{NFW} halo profile
\cite{Silk04}. 
Unfortunately, to create the $\Lambda$CDM relic density requires a cross-section of  $\le 1$pb.  
Nor  even for subdominant neutralinos can  SUSY
theory  easily produce such a large cross-section above 10 TeV.  A
spike may provide the best way of reconciling the high annihilation
flux with the dark matter relic abundance \cite{Silk04,Hooper04}.

{\it The redshift and mass of the first clouds, stars and IMBHs:}
The temperature $T$ of the first clouds that contain gravitationally bound
baryons depends mostly on the virial mass $M$ with
$M=10^6\msun \left[{T \over 10^3{\rm K}}{10 \over 1+z}\right]^{3/2}$, 
where we have scaled with typical values of the redshift $z$ and temperature $T$;  
cooling by $H_2$ requires a virial temperature in excess of several hundred degrees K.
The collapse redshift for clouds of mass $M$ is 
$\approx 0.6\nu \times 16 (M/10^6\msun)^{\epsilon},$
where $\epsilon \ll 1$ and $\nu$ is the height of the gaussian peak.
Hence the desired redshift is sensitive primarily to the peak rarity,
and must be in the range 10-30.  
The WMAP-inferred optical depth of the universe to Thomson scattering 
also suggests a redshift of 15-20 for the first objects that
reionised the universe.

The mass of the first objects can be estimated by noting that
the accretion rate is of order $v_s^3/G \sim 10^{-4}-10^{-3}\msun$/yr, 
where we take $v_s$ to be the virial velocity.  Accretion can only be halted by 
feedback, and it takes at least $10^6$ years to generate feedback from massive stars, 
hence the masses of the first stars are of order $100-1000\msun$, 
as found in numerical simulations of primordial gas clouds
by \cite{Abel02}.  Finally, in clouds of $10^4$K where 
Lyman alpha cooling dominates, the accretion
rate onto the protostellar core is around $0.03\msun$/yr, hence
feedback cannot inhibit the formation of an IMBH, of mass up to $\sim 10^4\msun.$

{\it Annihilation flux of (mini-)halo models with finite core and plateau:}
The (bolometric) annihilation flux integrated over a sphere of WIMPs of mass $m_\chi$ and number density 
$n(r)={\rho(r) \over m_\chi}$ at radius $r$ is 
\begin{equation}
L_{\chi\bar{\chi}}(<r) = \!\!\int_{\!0}^{\!r}\!\!\! 4\pi r^2 dr \left<\sigma v\right> m_\chi c^2 \left[{\rho(r) \over m_\chi},{\rho(r_p) \over m_\chi} \right]_{\min}^2 \!\!\!\!\!\!\!, 
\end{equation}
where annihilation imposes  a plateau of size $r<r_p$ on the present-day dark matter density because
neutralino number density above ${\rho(r_p) \over m_\chi} \equiv {H_0 \over \left<\sigma v\right>}
\sim {10^8\msun\pc^{-3} \over 50\GeV} \sim 10^8\cm^{-3}$ will annihilate away over a Hubble time
(e.g., \cite{GS99,UZK01}), where the cross-section $\left<\sigma v\right>/c \sim 1$pb, fixed by the relic abundance 
$\Omega_\chi$ required in a $\Lambda$CDM cosmology.
The total detectable flux quickly converges at large radius
since $\rho(r)$ generally drops as a steeper than $r^{-1.5}$ power-law beyond $r_p$.  
Adopting the typical SUSY mass $m_\chi=50\GeV$, we have
\begin{equation}
{L_{\chi\bar{\chi}}(<r) \over  2 \times 10^{3}\lsun} 
= \!\!\int_{\!0}^{\!r} \!\!\left[1,{\rho(r) \over 10^8\msun\pc^{-3}}\right]_{\min}^2 \!\!\!
{r^2  dr \over (10^{-3}\pc)^3}.
\end{equation}
Note that less than 10\% of the above calculated flux for neutralino models
will likely go into photons ($\gamma$-rays).  
More generally and crudely speaking, $L_{\chi\bar{\chi}}(<r) \propto m_\chi^{-1}$.

We model the density $\rho(r)$ of the minihalos and the MW halo  
with a subset of models of Zhao\cite{Zhao96} (also known as generalized-NFW profiles).
To be {\it conservative in estimating the annihilation signal},
we choose a subset of initial halo models {\it with a finite core and 
without any outer truncation}.  Our models are also  
inspired by the recent exponential-like model of Navarro et al. 
(\cite{N04}, N04 model hereafter), but  are 
{\it easier to work with analytically}.
Specifically, we model halos with density at radius $r$ given by 
\beq \label{eq:zhao}
\rho(r) = {\rho(r_c) \over 
\left[ x/2 + 1/2 \right]^{12} }, \qquad x \equiv \left({r \over r_c}\right)^{1 \over 3}. 
\eeq
Here $r_c$ is, as in an NFW model, where the local density power-law index $d\log \rho /d\log r = -2$.  
Our models are specified by two parameters $r_c$ and 
$\rho_c \equiv \rho(r_c)$, and the virial radius $r_{\rm vir} \equiv c r_c$ is given implicitly by
\beq
{3 M_{\rm vir} \over 4\pi r_{\rm vir}^3}
=176 \rho_{\rm crit}(z) 
={2 \rho(r_c) (1+ c^{1 \over 3}/5 + c^{2 \over 3}/55) \over (1/2+c^{1 \over 3}/2)^{11}},
\eeq
i.e., the mean density $\bar{\rho}(r_{\rm vir})=
{3 M_{\rm vir} \over 4\pi r_{\rm vir}^3}$ is approximately $176$ times the critical density 
$\rho_{\rm crit}(z)$ of the universe at a given formation redshift $z$.
Note the density ratio
$\bar{\rho}(r_c)/\rho(r_c) = 2.436$, $2.3178$ and $2.4644$ for our cored
model, NFW model\cite{NFW}, and N04 model.  Unlike
the NFW model, our model has an outer $r^{-4}$ density profile without
truncation, and a finite central density and a finite asymptotic mass.
For models with the same core radius and density nomalisation $\rho(r_c)$,
$M(r_{\rm vir})$ differs by only 10 percent between NFW models and our model for 
typical concentration $c=1-1000$. 
\begin{figure}
%\begin{center}
\resizebox{8cm}{!}{\includegraphics{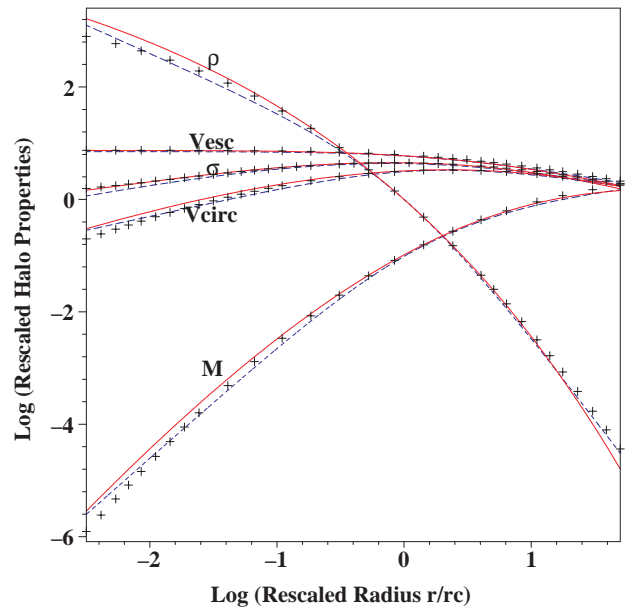}}
\caption{log-log radial profiles of the rescaled halo density $\rho(r)/\rho(r_c)$, escape velocity  
$V_{\rm esc}(r)/(K\sqrt{2})$, dispersion $2\sigma(r)/K$, circular velocity $V_{\rm circ}(r)/K$, and 
enclosed mass $M(r)/100\rho(r_c)r_c^3$ in our cored halo model (crosses),
NFW model (blue dashed lines) and N04 model (red lines).  
Here $K=\sqrt{G\rho(r_c)r_c^2}$, and $r_c$ is the core radius.
\label{fig1}}
%\end{center}
\end{figure}

There are a few advantages of our model.
It is straighforward to derive and show that the potential, 
the circular velocity curve and even the velocity dispersion curve of 
this class of halo models are all {\it simple analytical} rational functions of $r^{1/3}$,
while some special functions enter the NFW and N04 models.  
As shown in  Fig.1, our model reproduces the density,
mass, velocities of NFW and N04 very well over the range $0.01r_c < r < 100
r_c$ at least for systems with $c \ge 1$.  

{\it Formation of minihalos and massive black holes:}
Current numerical simulations suggest that  the first stars form from $H_2$ clouds
in dark minihalos of mass about $3\times 10^5 h^{-1}\msun$, virial radius of $~60h^{-1}\pc$
and virial velocity of $5\kms$, and initial 
virial temperature of $1000$ Kelvin
(for a molecular weight of unity) 
at a redshift of $z\sim 20$, when the universe was about $0.15$ Gyr old. 
 The cooling time is comparable to the dynamical time, 
both being of order 1 Myr,
and  self-gravitating isothermal clouds of order a thousand solar masses 
are able to develop.  Once the $H_2$ metal-free gas in these systems 
cools to a temperature of $\sim 200$K and density of $\sim 10^4\cm^{-3}$, 
the $H_2$ cooling saturates and subfragmentation 
ceases \cite{BrommLarson04}.
These clouds are generally not at the exact centres of the minihalos,
but have dynamically decayed to the central  parsec of the minihalos with
typical random motions of less than $10\kms$.  
The first stars are generally massive 
(about $100-1000\msun$) because of inefficient cooling with
zero metals.  They are short-lived.  According to \cite{HegerWoos02} (see their Fig.2), 
these very massive zero-metalicity stars have large helium cores,
which are unstable to pulsations due to electron-position pair instability.  
A supernovae leaving no remnant happens for $140-260\msun$ stars, and 
a more massive star collapses into an IMBH after lossing about 
half of the initial mass by wind.   However, Lyman alpha-cooling clouds would
be expected to form more massive IMBHs, up to $10^3-10^4\msun.$

There are several plausible outcomes for the IMBH and the gas inside a minihalo.  
In some cases, the IMBH could be ejected out of the shallow
potential well of the minihalo if born with a significant kick/recoil velocity
(greater than the $30\kms$ escape velocity at the centre of the minihalo)
\cite{Madau04}.
In other cases, the IMBH spirals to the centre of
the mini-halo due to dynamical friction (the timescale $t_{\rm df} \sim (r/\pc)^2
\sim 1$Myr for a thousand solar mass BH formed at 1pc).
At the centre or en route, these IMBHs  may accrete gas at a super-Eddington 
rate with a mass-doubling time scale $400\epsilon \ln(2) \approx 1-20$Myr
for a plausible range of the accretion efficiency $\epsilon=0.005-0.1$.  
Eventually after some uncertain period of order 10 Myr,
the first SNe progenitors are born in these minihalos, and shortly
afterwards (about 4 Myr) most of the gas in the minihalos (perhaps with the 
exception of the tightly bound accretion disk around the IMBH) is cleared out
when the SNe explode.  Thus the only remnants of 
the first generation of star formation are 
a population of minihalos and IMBHs, which are either well-centred 
due to dynamical friction or well-separated due to strong kicks.
To be specific, we consider minihalos specified by
eq.~(\ref{eq:zhao}) with $r_c=1.6\pc$ and $\rho(r_c)=500\msun\pc^{-3}$
(comparable to an NFW halo of $r_{\rm vir}=50\times r_c=80\pc$).
We assume conservatively that 10 percent of the minihalos harbour IMBHs 
which eventually grow to a final mass of $10^2\msun$ or $10^3\msun$.
These values are plausible for a r.m.s. kick velocity $\sim 100\kms$, 
and accretion efficiency $\epsilon \sim 0.02$ in the mini-halo in 10 Myrs.
So there are of order $10^3$ minihalos containing IMBHs.

{\it Adiabatic compression of dark matter in minihalos:}
Now consider the effect on the dark matter.  The formation and the growth of
the IMBH will compress the surrounding dark matter.  The compression is close
to adiabatic since the formation of the IMBH and the subsequent accretion are
on time scales of the cooling time and Salpeter time; both are somewhat
longer than the dynamical time.  Given the finite density and phase space
density of dark matter at an off-centre position, we expect a moderate
$r^{-3/2}$ power-law dark matter density distribution surrounding the IMBH.
This type of cusp (or "spike")  is the result of
growth in a uniform background\cite{BT}; a steeper cusp
occurs if a well-centred BH grows adiabatically (e.g., \cite{UZK01}).  
Such a mini-spike will survive dynamical friction, 
and be added to the central core of the minihalo once the IMBH
recenters.  We expect of order of $~m_{\bullet}\sim 1000\msun$ of dark matter
in this spike.  So the dark matter mass profile with a mini-spike is 
\beq 
M^{\rm sp}(r) = \left(1 + {b^{3/2} \over r^{3/2}}\right)M(r), \qquad M(b)=m_\bh,
\eeq 
where $b=0.045\pc$ or $0.14\pc$ so that $M^{\rm sp}(b)=2M(b)=m_\bh=10^2\msun$ or $10^3\msun$.

We consider several possibilities for the dark matter.  
Fig.2 shows the resulting dark halo density and the annihilation flux.  In the
simplest case, the minihalos have the pristine cored density profile;
the IMBH either never formed or has been kicked out of the system.  For a minihalo which keeps
its IMBH, the IMBH draws in  a mini-spike of dark matter with a plateau inside 
about 0.005pc depending on its mass.  The flux per log radius
bin is of order $10^4\lsun$, and most of it comes from within 0.01pc of the
minihalos.   
\begin{figure}
%\begin{center}
\resizebox{9cm}{!}{\includegraphics{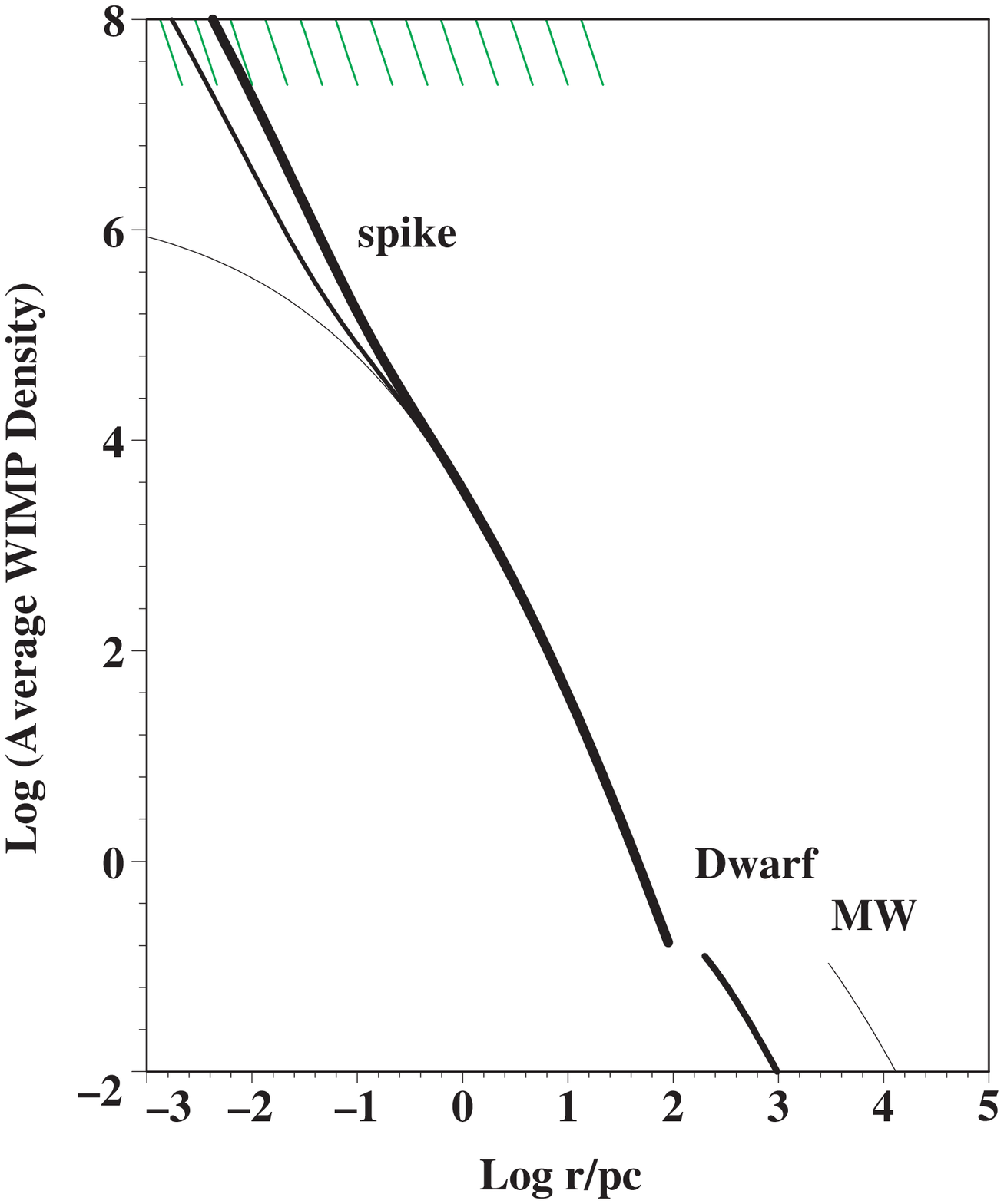}\includegraphics{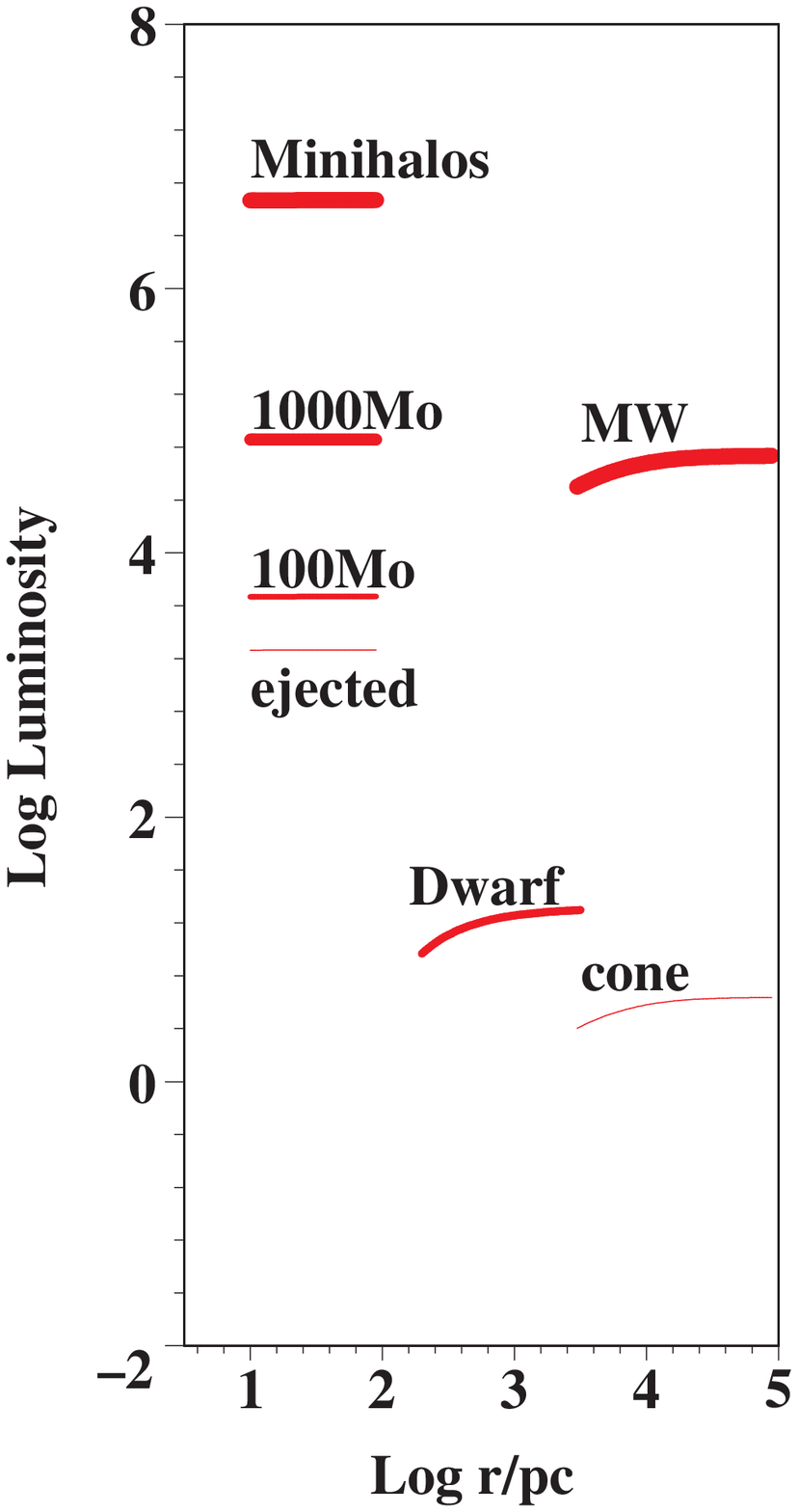}}
\caption{{\bf Left}: log-log plot of the radial profiles of spherically averaged density (black lines) 
in $\msun\pc^{-3}$ inside a mini-halo in three cases: 
a strong mini-spike around a $1000\msun$ IMBH (thick line),
or a moderate mini-spike around a $100\msun$ IMBH, or a pure cored minihalo
where the IMBH either was never formed or was ejected (thin line).
Lines (green parallel) of constant enclosed mass $1\msun, 10\msun, ..., 10^{12}\msun$ are shown
to guide the eye.  For comparison are the average densities  
inside a dwarf galaxy (lower thick line, with dispersion of $10\kms$ and a core of $200\pc$) 
and inside the pristine MW halo (lower right thin line).
{\bf Right}: the logarithm of the rescaled annihilation (bolometric) luminosity 
${L_{\chi\bar{\chi}}(<r) \over \lsun} \times {m_\chi \over 50\GeV}$
(nearly horizontal red lines to the right) of these models. 
Also shown are the total luminosity of $10^3$ minihalos with $100\msun$ IMBHs
(thickest black at top middle), the luminosity of the MW as a whole (thick red)
or scaled down by a factor ${\Delta \Omega \over 4\pi}$ for
a cone of $\Omega=10^{-3}$ (thin red).  The minihalos
stand out as clumps above the MW background. 
\label{fig2}}
%\end{center}
\end{figure}

{\it Observability of minihalos inside the MW:}
By a redshift of $z=1-6$, nearly all the minihalos and IMBHs are
bound inside galaxy-sized objects.  About $10^3-10^4$ should be bound
inside a Milky-Way-sized galaxy \cite{MadauRees01}, which we assume to have
a density given by eq.~(\ref{eq:zhao}) with $r_c=11\kpc$ and 
$\rho(r_c)=0.0055\msun\pc^{-3}$ (this is consistent with the local DM density
and resembles an NFW model of $r_{\rm vir}=8 r_c$).
These minihalos should be
distributed in a way similar to the overall dark matter profile of the
galaxy, producing a clumpy annihilation brightness distribution; 
we assume the stars and the central 
SMBH in the MW form later.

A typical nearby minihalo at about 3 kpc from the Sun sustains a solid angle 
$\Delta \Omega \sim 10^{-3}$ (about the resolution of EGRET) comparable to the 
angular size of dwarf galaxy satellites at 100kpc, but minihalos   
are much more luminous and brighter in annihilation than dwarf galaxies
because of minihalo's much denser core than dwarf galaxies; the latter have 
central dark matter densities of $(0.05-0.5)\msun\pc^{-3}$ \cite{MateoReview}.
Except towards the Galactic center a minihalo stands out in density and annihilation flux 
from the MW background in a beam of $\Delta \Omega \le 10^{-3}$ 
at all distances (cf. Fig.~2). 
A population of $10^3$ minihalos with $100\msun$ IMBHs
is about 100 times more luminous than the MW background as a whole.
These minihalos are low-mass ($\sim 10^6\msun$) but dense 
(about $1\msun\pc^{-3}$ on a scale of 30pc) 
enough so that 99 percent of the minihalos survive intact in the halo, and follow the
overall distribution of dark matter.
There are of order 10 minihalos within 1 kpc of the MW centre, where
dynamical friction could bring one of these minihalos to the centre of the MW 
within a Hubble time (cf. Fig.3 of \cite{UZK01}), without being slowed down
by tidal stripping \cite{Zhao04}; note that these minihalos are 
typically denser than the MW halo on a scale of 60pc.
If the IMBH of the minihalo could serve
as a seed for the supermassive BH in the MW.

The isolated IMBHs are virtually invisible.  The X-ray luminosity due to 
Bondi-Hoyle acretion of an IMBH passing through the MW gaseous disk is
estimated to be $(\epsilon/0.1) (n/30\cm^{-3}) (m_\bh/300\msun)^2 \lsun$
\cite{LO85}.  So the signal is much fainter than the 
million-solar-mass BHs of Lacey \& Ostriker\cite{LO85}.  
Such a low luminosity is just detectable with
present facilities if the IMBH is  close-by (say 2 kpc), but the chance of
a BH passing through a cold H$_2$ or HI cloud (with filling factors 0.3\%
and 2\% respectively) at a given time is very low.  The local density
of IMBHs is about $0.3\kpc^{-3}$ assuming $10^4$ IMBHs in the whole halo,
and we expect only of order one IMBH hitting a cloud with a filling factor
of $\sim$1\% in the entire disk of 8 kpc in radius and 200pc in thickness,
an approximation for the cold H$_2$ and HI cloud distribution. 
Also being much fewer than the
million-solar-mass BHs of\cite{LO85}, the minihalos 
do no damage to the MW disk.  

Despite being a dynamically subdominant population, 
the minihalos and IMBHs speed up two-body relaxation in the
MW.  The relaxation time is given by\cite{LO85}
${ 1{\rm Myr} \over t_{\rm rlx}} 
= {\rho(r)\sigma^{-3} \over 10^6\msun\pc^{-3}(100\!\kms)^{-3} }
{\bar{M} \over 10^3 \msun }$, 
where $\bar{M}={f_M \ln \Lambda \over 20}{M^2 + m_\bh^2 \over M + m_\bh } 
\sim 1-1000\msun$ is the mass density-weighted mean
mass of deflectors depending on tidal stripping; the lower value is
for minihalos well inside 1-10 kpc of the Galactic centre, and 
$f_M\sim 10^{-3}$ is the fraction of matter mass density in the form of minihalos
of mass $M$.  The heavier the deflectors, the faster the
relaxation.  So compared to the case of  purely stellar deflectors,
the relaxation is about $\bar{M}/\msun \sim 1-1000$ times more efficient.  
This might help to refill the loss-cone of the 
central BH by scattering stars and dark matter particles\cite{ZHR02}.
To refill the loss-cone of $3\times 10^6\msun$ BH on the scale of the $1\pc$ 
would require a relaxation time of order a Hubble time.

Our model predicts around 1500 $\lsun$ (bolometric) per solar mass in spike
neutralinos.  Observationally, the annihilation signal is best seen via gamma rays.  EGRET
observations set upper limits on unidentified sources, in particular on the source near 
the Galactic Centre, in gamma rays above 1 Gev of around $100\lsun$ \cite{Nolan03}. 
For our models to be consistent with observational data in the MW 
and a gamma ray branching ratio of order 10 percent, 
we require of order a solar mass of the neutralinos in the spike 
for typical SUSY neutralinos with $m_\chi=50\GeV$.  
This is consistent with our models of adiabatic growth of $10^2\msun$ IMBHs 
and models with ejected IMBH.  The data would also be consistent with 
the $10^3\msun$ IMBH model in case of heavier neutralinos 
($m_\chi \sim 10{\rm Tev}$ \cite{BertMerr05}).
Evidently the GLAST satellite,
with two orders of magnitude more sensitivity than EGRET, will be capable of
setting stringent constraints on halo IMBH.

We thank the referee for many detailed comments which help to 
improve the presentation here.  HSZ acknowledges travel and publication support from 
Chinese NSF grant 10428308 and UK PPARC Advanced Fellowship.

%\vfill\eject

%%%%%%%%%%%%%%%%%%%%%%%%%%%%%%%%

%\bsp

\label{lastpage}

\end{document}